%% file: Tex_files/main.tex
\documentclass[10pt, conference]{IEEEtran}
\usepackage[english]{babel}
\newtheorem{proposition}{Proposition}
\usepackage[
  letterpaper,
  top=0.75in,
  bottom=1.022in,
  left = 0.625in,
  right=0.625in
]{geometry}
\usepackage{amsmath}
\usepackage{epsfig}
\usepackage{amssymb}
\usepackage{graphicx}
\usepackage{blindtext}
\usepackage{multicol}
\usepackage{lipsum}
\usepackage{subcaption}
\usepackage[bookmarks=false]{hyperref}
\usepackage{pdfpages}
\usepackage{array, makecell} %
\usepackage{dcolumn}
\usepackage{algorithm,algorithmic}
\usepackage{cancel}
\hypersetup{
    colorlinks=true,
    linkcolor=blue,
    filecolor=magenta,      
    urlcolor=cyan,
    pdftitle={Overleaf Example},
    pdfpagemode=FullScreen,
    }
\title{Asymmetric Modulation Design for Fluid-Antenna SWIPT Systems}
\author{
\IEEEauthorblockN {
   Ahsan Mehmood$^{\dagger}$,
   Ioannis Krikidis$^{*}$, Fellow, IEEE        %%% instead of \IEEEauthorrefmark{1}
   and Ghassan M. Kraidy$^{\dagger}$, Senior Member, IEEE %%% instead of \IEEEauthorrefmark{1}\IEEEauthorrefmark{4}
}\\
\IEEEauthorblockA {
 $^{\dagger}$Department of Electronic Systems, Norwegian University of Science and Technology, Norway\\
$^{*}$Department of Electrical and Computer Engineering, University of Cyprus, Cyprus}
\IEEEauthorblockA {
Email: {\{ahsan.mehmood, ghassan.kraidy\}}@ntnu.no, krikidis.ioannis@ucy.ac.cy
}
}

\begin{document}
\maketitle
\input{Tex_files/Abstract}
\input{Tex_files/Intro}
\input{Tex_files/System_model}
\input{Tex_files/Appendix_A}

\bibliographystyle{ieeetr}
\bibliography{main}
\end{document}

%% file: Tex_files/Abstract.tex
\begin{abstract}
   In this work, we propose the design of modulation schemes that improve the rate-energy region of fluid antenna-assisted  simultaneous wireless information and power transfer (SWIPT) systems. By considering the nonlinear characteristics of practical energy harvesting circuits, we formulate a dual-objective rate–energy (RE) region optimization problem to jointly maximize the discrete-input mutual information (DIMI) and harvested current.
   The problem is solved using the $\epsilon$-constraint method and optimized constellations are designed for various energy harvesting thresholds. {We then evaluate the performance of the optimized constellations under three different fluid antenna (FA) port selection strategies: (i) Best Port, (ii) Fixed Port, and (iii) Random Port.} Our simulation results demonstrate significant performance gains of optimized constellations over conventional constellations in both information rate and energy harvesting.
\end{abstract}


%% file: Tex_files/Intro.tex
\section{Introduction}
Sixth-generation (6G) networks aim to support massive low-power devices in applications like internet of things, sensor networks, and smart grids, but their scalability is hindered by the limitations of conventional power sources, such as batteries requiring frequent recharging and grid electricity being inaccessible in remote areas. To mitigate this drawback, simultaneous information and power transfer (SWIPT) emerges as a viable solution that enables the large-scale deployment of wireless-powered (WP) devices without reliance on traditional energy sources \cite{ mehmood2022throughput}. However, these WP devices operate under strict energy constraints, limiting the number of radio frequency (RF) components that they can support in their hardware resulting in degradation of the reliability of communication. 

In traditional communication systems, multiple-input multiple-output (MIMO) technology has been employed to improve the  reliability of transmission by exploiting the spatial diversity they provide \cite{10753482}. However, their high energy and computational demands make it impractical for WP devices. In this context, fluid antenna (FA) systems have emerged as a low-power alternative, offering MIMO-like gains with minimal hardware through software-controlled movement, flexible switching, and reconfigurable positioning of liquid metal ports \cite{wong2022bruce}. By dynamically adapting to channel conditions and efficiently leveraging spatial diversity \cite{10753482}, FA-assisted receivers enhance the performance of SWIPT systems while reducing RF chains requirements, making them ideal for energy-constrained SWIPT systems.

SWIPT aims to enable energy-efficient operation by simultaneously maximizing both energy harvesting and information transfer. It employs the same signal constellation for both purposes \cite{clerckx2018fundamentals}, which necessitates careful signal constellation design to optimally balance the dual objectives. However, research on constellation design for enhancing both {energy harvesting (EH)} and data transmission remains limited.
Existing research on constellation design for power-splitting (PS) SWIPT has largely focused on evaluating conventional modulation schemes, with limited efforts toward optimizing constellations for enhanced EH and data transmission. The study in \cite{7511404} analyzed the EH performance of standard modulation schemes such as M-ary phase shift-keying (M-PSK), M-ary pulse amplitude modulation (M-PAM), and M-ary quadrature amplitude modulation (M-QAM), revealing that signals with a higher peak-to-average power ratio (PAPR) enhance EH efficiency. 

Building on this, \cite{9860473} introduced spike-QAM, a high-PAPR variant that enhances EH by positioning selected constellation points farther from the origin, while \cite{9593249} proposed circular QAM, optimized for both EH and symbol error rate (SER) performance. However, these studies relied on linear EH models, which, though simple, fail to accurately capture practical EH circuit behavior.
\begin{figure*}
    \centering
    \includegraphics[width=0.8\linewidth]{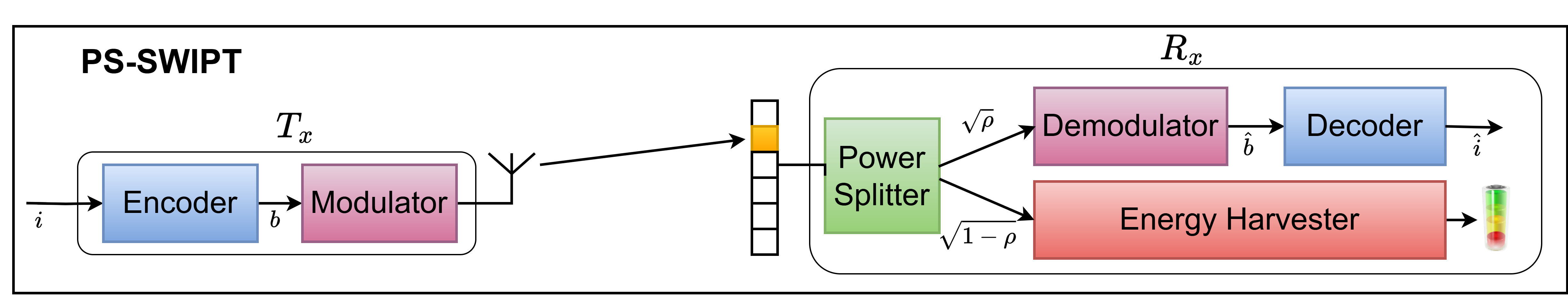}
    \caption{FA-assisted PS-SWIPT communication system.}
    \label{fig:2}
\end{figure*}
%\vspace{-1mm}
To address this issue, \cite{7264986} and \cite{7547357} introduced nonlinear models based on a sigmoidal function and a Taylor series expansion of diode characteristics, respectively. It was demonstrated in \cite{7547357} that EH efficiency depends not only on the PAPR but also on the phase range of constellation points. Building on this, \cite{varasteh2019learning} developed a neural network-based encoder-decoder framework to design modulation schemes using both sigmoidal and Taylor-series-based EH models, though without explicitly accounting for the phase range in the optimization. 

Based on this, authors of \cite{bayguzina2019asymmetric} introduced asymmetric PSK (APSK) to expand the rate–energy tradeoff region. However, despite APSK’s enhanced EH efficiency, its constrained Euclidean distance leads to degraded decoding performance at higher modulation orders.
By considering both constrained phases and amplitudes of modulated symbols, \cite{9951152} proposed coded asymmetric  QAM (AQAM), achieving improved EH and information decoding performance. These constellation designs did not consider the fundamental SWIPT performance criterion that is PAPR, thus highlighting the need for more optimized asymmetric modulation schemes in SWIPT systems.

{To this end, we propose a novel approach for designing an asymmetric modulation scheme that jointly optimizes all key constellation parameters, including the PAPR. The main contributions of this work are summarized as follows:
\begin{itemize}
\item \textit{Integration of the discrete-input mutual information (DIMI) and a practical EH model:} We adopt a practical nonlinear EH model that implicitly accounts for the effects of both PAPR and the phase range of AQAM, following \cite{bayguzina2019asymmetric, 9951152}. This enables an accurate characterization of harvested energy. In parallel, we employ the DIMI as a performance metric to quantify the achievable rate, in contrast to the conventional Shannon capacity which assumes a continuous Guasian input.
\item \textit{Derivation of a tractable lower bound on the DIMI:} We establish a computationally efficient lower bound on DIMI, expressed in terms of the best FA port channel gain. This bound serves as the optimization objective and facilitates efficient solution of the rate–energy (RE) region optimization problem.  
\item \textit{Formulation of the RE-region optimization problem:} We introduce a novel dual-objective RE-region optimization framework that fully captures the geometric structure of the constellation. Specifically, the optimization jointly considers the phase, magnitude, and phase range of the constellation points, thereby enabling a fine-grained characterization of the rate–current tradeoff.\footnote{Following \cite{9377479}, harvested current is used as a proxy for harvested energy; thus, the RE region specifically refers to a rate–current region.}  
\item \textit{Solution via $\epsilon$-constraint method:} We solve the RE-region optimization problem using the $\epsilon$-constraint method, which enables construction of the complete Pareto front. This approach provides a comprehensive characterization of the tradeoff between achievable rate and harvested energy.  
\end{itemize}
We solve the RE-region optimization problem for a range of EH thresholds, thereby designing asymmetric constellations tailored to diverse information transfer and energy harvesting requirements. Finally, the performance of the optimized constellations is evaluated under various FA port selection strategies.}

{The remainder of this paper is organized as follows. Section~\ref{Sys_model} introduces the considered system and channel models along with the practical non-linear energy harvesting model adopted in this work. Section~\ref{opt_model} formulates the RE-optimization problem and outlines the proposed optimization approach. Section~\ref{perf_eval} describes the simulation setup in detail and presents a comprehensive discussion of the obtained results. Finally, Section~\ref{Conclusion} summarize the key finding and discuss future research direction.}

%% file: Tex_files/System_model.tex
\section{System Model}\label{Sys_model}
We consider digital transmission over a block-fading channel with PS-SWIPT, in which the transmitter is equipped with a {conventional} antenna, and the receiver is equipped with an $N$-port FA defining the $N$ fading blocks, as illustrated in Fig. \ref{fig:2}. \\
In the PS-SWIPT configuration, a fraction $1- \rho$ of the received
signal’s power is used to harvest energy, while the remaining
$\rho$ is used for information transfer.  Thus, the baseband channel model is given by
% \vspace{-1mm}
\begin{equation}
% \vspace{-1mm}
    y = h_n x + \eta,
\end{equation}
where $\eta$ is an {independent identically distributed} (i.i.d.) circularly symmetric complex Gaussian
noise of zero mean and variance $N_0$, $x \in \mathbb{C}$ is the modulated symbol, and $y\in \mathbb{C}$ is the received symbol. %{ and a fraction of it (e.g., $y_I = \sqrt{\rho}(h_n x + \eta)$) represents the signal used for information decoding}.
\\
\indent We consider a correlated fading model between the $N$ ports of the FA, where the selected fading $h_n$ (corresponding to port $n$) is the $n^{th}$ 
element of vector $\mathbf{h} = \mathbf{g\sqrt{C}}$, with $\mathbf{g}$ being a length-$N$ vector of zero-mean unit-variance complex Gaussian random variables, and $\mathbf{C}$ being the correlation matrix whose entries are given as $\mathbf{C_{i,j}} = J_o \left(2\pi(i-j)\frac{W}{N-1}\right)$, where $J_o (\cdot)$ denotes the zero-order Bessel function of the first kind, and $W$ represents the FA length in terms of wavelength \cite{10753482}. The results in this paper are valid whatever the choice of $h_n$ is, however optimal performance is obtained when the best port ({\em i.e.,} the one with largest fading magnitude) is selected. We assume that the channel coefficients are known to the receiver, but not to the transmitter.\\
\indent To take into account the nonlinearity of the rectification process, we consider a nonlinear energy harvesting model \cite{bayguzina2019asymmetric}.
{The signal at the input of the  energy harvester can be expressed as
    $y_E = \sqrt{1-\rho}(h_n x + \eta)$,
where} power can expressed, by neglecting the effect of noise as $\mathrm{P_{R}} \approx  (1-\rho)|{h}_nx|^2$.
{Hence, the diode direct current (DC) $\textit{i}_{DC}$ at the output of the rectenna circuit—which determines the amount of harvested energy—can be expressed as in \cite{bayguzina2019asymmetric} and is given in (\ref{Eq:i-h}). In this expression, the diode parameters are given by $\mathrm{R_s} = 1~\Omega$, $\mathrm{k_2} = 0.0034$, and $\mathrm{k_4} = 0.3829$, and the phase of the modulated symbols is constrained within the interval $[-\delta, \delta]$. }

\begin{algorithm*}
\begin{equation}\label{Eq:i-h}
      \textit{i}_{DC}(h_n) = k_o + k_2R_s(1-\rho)\mathbb{E}_X\{|h_n x|^2\}+\frac{3}{4}k_4R_s^2(1-\rho)^2\mathbb{E}_X\{|h_n x|^4\}e^{-\frac{2}{3}\delta}
\end{equation}

 \begin{equation}\label{Eq:PS_MI}
      I(X;Y)_{h_n} = \log_2(M) - \frac{1}{M}\sum_{k=1}^M\mathbb{E}_{N} \biggl\{ \log_2\sum_{i=1}^M \exp\left(-\frac{|\sqrt{\rho}h_n x_k+\eta-\sqrt{\rho}h_n x_i|^2-|\eta|^2}{2\rho N_0}\right)  \biggl\}
\end{equation}
\end{algorithm*}
\section{Asymmetric QAM  design based on rate-energy regions}\label{opt_model}
In this section, we investigate the design of asymmetric QAM \cite{bayguzina2019asymmetric} that allow to optimize the RE-region of PS-SWIPT FA channels. In terms of rate, we consider the DIMI \cite{1056454}, which gives the fundamental limit of discrete modulations. The DIMI $I(X;Y)_{h_n}$ of the FA channel with port $n$ activated is given in (\ref{Eq:PS_MI}), where $M$ is the modulation size, and $\mathbb{E}_N$ denotes expectation over noise. The RE region for PS-SWIPT FA systems can now be defined as
\begin{equation}
% \vspace{-1mm}
    RE = \left\{ (\mathcal{R}, \mathcal{I}): \mathcal{R} \leq \mathbb{E}\{I\}, \mathcal{I} \leq \mathbb{E}\{\textit{i}_{DC}\} \right\}.
\end{equation}

With this definition the RE region optimization problem can be expressed as a dual-objective optimization problem as,
% \subsection{\cancel{RE Region Optimization }}
\begin{subequations}
\label{algo_main}
\begin{align}
\mathcal{P}{^S_1:} \: 
\underset{\delta, \mathbf{r}, \mathbf{\theta}}{\text{maximize}}  \quad &  \mathbb{E}_{h_n} \{I(X;Y)_{h_n}\}\label{13a} \\
\underset{\delta, \mathbf{r}, \mathbf{\theta}}{\text{maximize}} \quad  & \mathbb{E}_{h_n}\{\textit{i}_{DC} (h_n)\} \label{13b} \\
\text{subject to:} \quad 
&0 \leq \delta \leq \pi/2\label{10g}\\
& -\delta \leq \theta_k \leq \delta \ \forall k = 1,...,M\\ 
& \max\{\theta_1,\theta_2,...,\theta_M  \} = \delta\label{13c}\\
& \min\{\theta_1,\theta_2,...,\theta_M  \} = -\delta\label{13d}\\
&  \frac{1}{M}\sum_{k = 1}^M |r_ke^{j\theta_k}|^2 = 1\label{13e}\\
&\frac{\max_k (|r_ke^{j\theta_k}|^2)}{\frac{1}{M}\sum_{k=1}^M  |r_ke^{j\theta_k}|^2} \leq PAPR_{max}\label{13f}
\end{align}
\end{subequations}

This RE-region optimization problem aims to maximize both the average mutual information and the  average harvested energy by optimal geometric shaping of the constellations. Note that, in $\mathcal{P}{^S_1}$, the constellation points are expressed in their polar representation as $x_k=r_k e^{j \theta_k}$ with $r_k$ and $\theta_k$ representing the amplitude and the phase of $k^{th}${,} constellation point respectively. Moreover, $\boldsymbol{\theta} = [\theta_1, \theta_2, \ldots, \theta_k, \ldots, \theta_M]$ and $\mathbf{r} = [r_1, r_2, \ldots, r_k, \ldots, r_M]$ represent the phase and magnitude vectors of the constellation, respectively. 
The constraints (\ref{10g})-(\ref{13d}) ensure that the phase of each constellation point is confined within $[-\delta,\delta]$, which allows to exerce control on the last term in (\ref{Eq:i-h}). 
Moreover, the constraints (\ref{13e}) and (\ref{13f}) ensure that the average energy of constellation points is unity and the PAPR of the optimized constellation scheme is below the desired maximum PAPR ($PAPR_{\max}$).

\begin{figure*}
\label{Fig2}
\centering
\begin{subfigure}{0.24\textwidth}
\includegraphics[width= 1\linewidth]{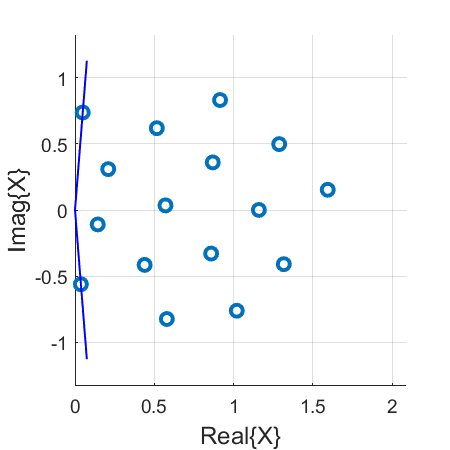} 
\caption{$\epsilon = 0.08$}
\label{fig:subim1}
\end{subfigure}
\hfill
\begin{subfigure}{0.24\textwidth}
\includegraphics[width=1\linewidth]{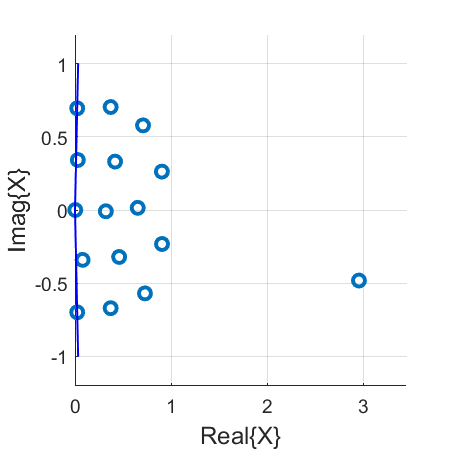}
\caption{$\epsilon = 0.30$}
\label{fig:subim2}
\end{subfigure}
\hfill
\begin{subfigure}{0.24\textwidth}
\includegraphics[width=1\linewidth]{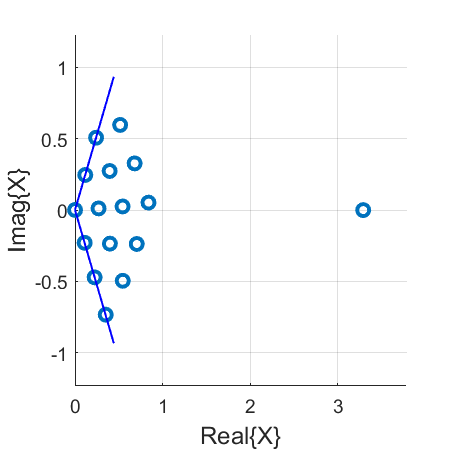}
\caption{$\epsilon = 0.51$}
\label{fig:subim3}
\end{subfigure}
\hfill
\begin{subfigure}{0.24\textwidth}
\includegraphics[width= 1\linewidth]{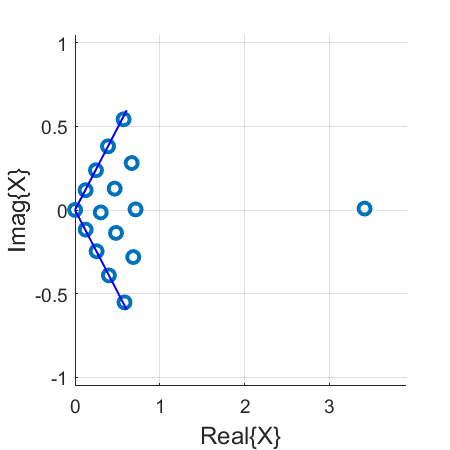} 
\caption{$\epsilon = 0.72$}
\label{fig:subim4}
\end{subfigure}

\caption{Optimized AQAM constellations of size $M = 16$ for different values of $\epsilon$ with $\gamma_d = 17 dB$, $W = 0.05$ and $N = 100$. }
\label{fig:image2}
\end{figure*}

The RE-region optimization problem can be solved by converting it into a single objective optimization problem by using the $\epsilon$-constraint method \cite{9377479}. Using this method, one of the two objective functions, namely (\ref{13b}), is converted into a constraint. However, to solve this single objective optimization problem with an $\epsilon$-constraint, an analytical expression of the expectation in $\mathbb{E}_{h_n} \{I(X;Y)_{h_n}\}$ is needed, which is difficult to obtain. This  expectation can thus be only evaluated through Monte Carlo simulations. However, an objective function that requires Monte Carlo simulation is computationally expensive for an optimization algorithm, which needs to evaluate the objective function many times to find an optimal solution. To this end, we construct an easy-to-optimize approximate lower bound of (\ref{13a}), which does not require Monte Carlo simulations, as presented in the following proposition.
\begin{proposition}
A lower bound on the objective function defined in (\ref{13a}) can be computed as
\begin{equation}\label{Eq:PS_MI_LB_main}
            \begin{split}
            \tilde{I}_{LB}(X;Y) \approx  m - \frac{1}{M}\sum_{k=1}^M  \log_2\sum_{i=1}^M
            e^{-\frac{\rho \mathbb{E}_{h_n}\{|h_n|^2\}}{2 N_o}|x_k-x_i|^2 }.
            \end{split}
        \end{equation}
        where $m=\log_2(M)$ denotes the number of bits per modulated symbol.\\
\end{proposition}
\textit{ Proof}: See the proof in Appendix \ref{Prop1}.\\

With this computationally efficient objective function, the rate-energy optimization problem can be expressed as,
\begin{subequations}
\begin{align}
\mathcal{P}{^S_2:} \: 
\underset{\delta, \mathbf{r}, \mathbf{\theta} }{\text{maximize}}  \quad &  \tilde{I}_{LB}(X;Y)\label{lb}\\
\text{subject to:} \quad & \mathbb{E}_{h_n}\{\textit{i}_{DC} (h_n)\} \geq \epsilon \label{11b} \\
& (\ref{10g})-(\ref{13f}),
\end{align}
\end{subequations}
{where the objective function} in (\ref{13a}) is replaced by its lower bound (\ref{Eq:PS_MI_LB_main}) in (\ref{lb}) and the function in (\ref{13b}) is replaced by the $\epsilon$-constraint (\ref{11b}). Note that $\epsilon$ is a lower limit on the average harvested current and the tradeoff between the two objectives of $\mathcal{P}{^S_1}$ can be obtained by varying $\epsilon$ between its lower and the upper limit (e.g. $\epsilon_{min}\leq \epsilon \leq \epsilon_{max}$) in $\mathcal{P}{^S_2}$. The lower limit on $\epsilon$ is obvious and can be set as $\epsilon_{min} = 0$, whereas $\epsilon_{max}$ above which $\mathcal{P}{^S_2}$ does not remain feasible can be obtained by maximizing (\ref{13b}) under the constraints (\ref{13e})-(\ref{13f}).

\section{Simulation Results}\label{perf_eval}
In this section, we present the simulation setup, describe the optimized constellations, and evaluate their performance using multiple performance metrics in an FA-assisted SWIPT system.

\subsection{Optimized Geometric Shaping}
{To find the optimized constellation points, we set the design parameters as $PAPR_{\text{max}} = 15$ and $\rho = 0.5$ in (\ref{11b}) to solve the optimization problem. Evaluating the objective function (\ref{Eq:PS_MI_LB_main}) {within the optimization process requires computing $\gamma = \mathbb{E}\{|h_n|^2\}\gamma_d$, which corresponds to the average fading signal-to-noise ratio (SNR), with the design SNR being $\gamma_d = \rho/2N_o$.} This average fading SNR can be pre-calculated if the channel statistics are perfectly known, or alternatively, set to a desired target level. In practical scenarios, however, an approximate value of $\gamma_d$ can be estimated by averaging over a large number of channel realizations.}
\begin{figure}
\begin{center}
\includegraphics[width=0.8\linewidth]{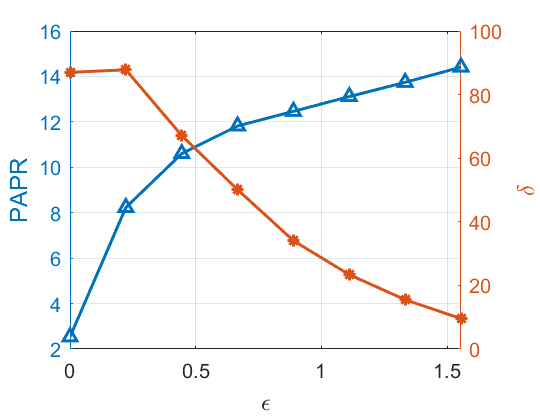}
\caption{Impact of $\epsilon$ on PAPR and phase range $\delta$ of optimized AQAM schemes. The simulation parameters are $M = 16$, $N = 100$, $W = 0.5$, $\rho = 0.5.$}
\label{fig:delta_papr}
\end{center}
\end{figure}

{In our implementation, we generate $L = 10^6$ correlated channel samples across $N = 100$ closely spaced ports of the fluid antenna, resulting in a set of channel realizations represented by the matrix $\mathbf{H} \in \mathbb{C}^{L \times N}$. {For each realization $l$, the best port is selected based on the channel gain, i.e., {$h_b^{(l)} = \mathbf{h}^{(l)}(b)$}, where $\mathbf{h}^{(l)}$ denotes the $l^{\text{th}}$ row of the channel matrix $\mathbf{H}$, and $b$ is the index of the port corresponding to the maximum channel gain, i.e., $b = \arg\max_{i} |\mathbf{h}^{(l)}(i)|$.}
 The resulting sequence of best-port channel gains is collected in the vector $\mathbf{h}_b = [h_b^{(1)}, h_b^{(2)}, \dots, h_b^{(L)}]$, which is then used to compute the expectation over these best port channel realization and consequently to compute $\gamma_d$.}
% Here it is worth noting that although our modulation schemes are designed using the single best-port FA channel, the proposed framework is general and can be extended to other channel models, including conventional SISO, multi-port FA, and MIMO systems
With this pre-calculated $\gamma_d$,  we solve the proposed optimization problem $\mathcal{P}{^S_2}$ with global search approach using interior-point method \cite{boyd2004convex} (``fmincon'' function in MATLAB). This optimization problem $\mathcal{P}{^S_2}$ is optimized with respect to the phase range $\delta$ and constellation points $\{x_k\}_{k=1}^M$ for a range of $\epsilon$ values to get different constellation schemes and to get insights into the rate-energy tradeoffs. Note that the constellation points in $\mathcal{P}{^S_2}$ are represented in polar form (e.g., $x_k = r_ke^{j\theta_k}$) and the phase and the amplitude of each constellation point serve as the optimization variables. 

For each value of $\epsilon$ the optimization problem is solved and the resulting geometrically shaped constellations are obtain. {Specifically, we solve the optimization problem $\mathcal{P}{^S_2}$ with $PAPR_{\max} = 15$, $\rho = 0.5$, $N_o = \rho/(2\gamma_d)$, $\gamma_d = 17dB$ and with eight equally spaced values of $\epsilon$ in the range [0.08, 1.57] to obtain eight geometrically shaped constellations for modulation order $M=16$.} Optimized constellations are shown for different values of $\epsilon$ in Fig. \ref{fig:subim1} - Fig. \ref{fig:subim4}.

{It is noteworthy that as $\epsilon$ increases, the PAPR increases while the phase range $\delta$ decreases. This is intuitive because the {dc current $\textit{i}_{DC}$}, which must be greater than $\epsilon$, is inversely proportional to the phase range $\delta$ and directly proportional to $\mathbb{E}_X\{|x|^4\}$ {(see (\ref{Eq:i-h}))}. Since the PAPR is directly proportional to $\mathbb{E}_X\{|x|^4\}$, an increase in $\mathbb{E}_X\{|x|^4\}$ also leads to an increase in  PAPR.} Therefore, increasing $\epsilon$ forces the optimization problem to shape the geometry of the constellations in such a way that the PAPR increases and the phase range decreases so as to satisfy  the constraint  (\ref{11b}). This behavior is further highlighted in Fig. \ref{fig:delta_papr}, in that the PAPR increases together with $\epsilon$, while the angle decreases with increasing $\epsilon$.
\begin{figure}
    \centering     \includegraphics[width=0.8\linewidth]{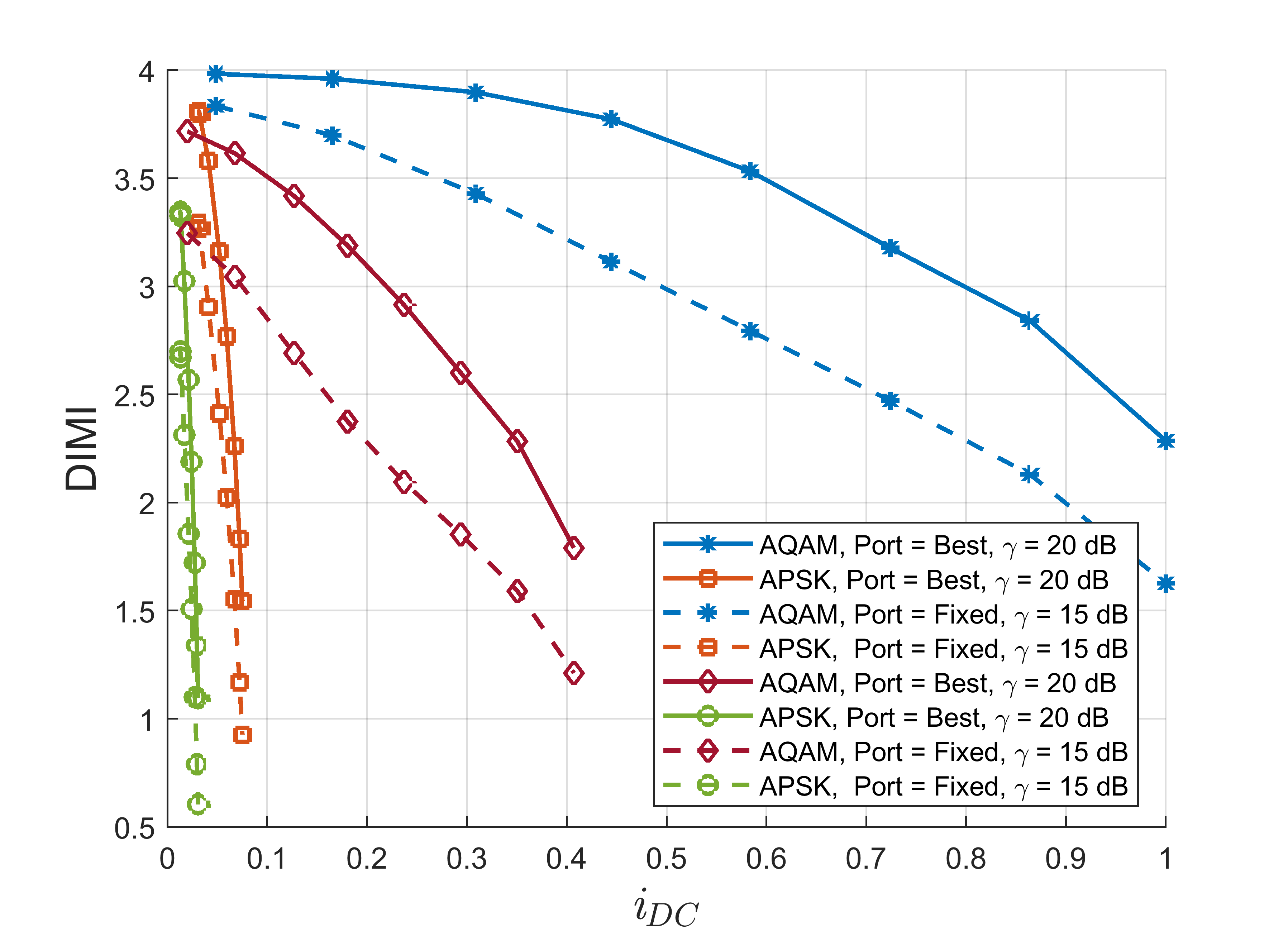}
 \caption{Impact of AQAM optimization on the rate-energy regions. The simulation parameters are $M = 16$.}
    \label{fig:rate-energy-region}
\end{figure}

\subsection{Performance Evaluation}
{We now evaluate the performance of the optimized constellation schemes in the FA-assisted SWIPT system. Specifically, we use three performance metrics: i) RE-region presented as a tradeoff between $\textit{i}_{DC}$ and the DIMI ii) DIMI versus SNR iii) $\textit{i}_{DC}$-SSR tradeoffs.} We evaluate the system performance under three port selection strategies:
    \textbf{i)} {Best port: $h_b^{(l)} = \mathbf{h}^{(l)}(b)$}, where $b$ is the index of the port corresponding to the maximum channel gain, i.e., $b = \arg\max_{i} |\mathbf{h}^{(l)}(i)|$.
    \textbf{ii)} Random port: $h_r^{(l)} = \mathbf{h}^{(l)}(r)$, where $r$ is a random integer uniformly drawn from the set $\{1, 2, \dots, N\}$.
    \textbf{iii)} Fixed port: $h_f^{(l)} = \mathbf{h}^{(l)}(f)$, where $\mathbf{h}^{(l)}$ is $l^{th}$ column of $\mathbf{H}$ and $f$ is a predetermined (fixed) port index.
{To assess the performance of these optimized constellations, we again generate a channel matrix $\mathbf{H}$ with $10^6$ correlated channel samples across $N = 100$ closely spaced ports of the fluid antenna. From this channel matrix, the best port, fixed port, and random port  channel realizations are obtained as $\mathbf{h}_b = [h^{(1)}_b, h^{(2)}_b, ..., h^{(10^6)}_b]$, $\mathbf{h}_f = [h^{(1)}_f, h^{(2)}_f, ..., h^{(10^6)}_f]$, and $\mathbf{h}_r = [h^{(1)}_r, h^{(2)}_r, ..., h^{(10^6)}_r]$ respectively. We also generate an equal number of noise samples to evaluate the performance metrics.}

{To generate the RE-regions, we evaluate the average DIMI and the average harvested current $\text{i}_{DC}$ for each AQAM optimized for a given energy harvesting threshold $\epsilon$. For each fluid antenna port selection strategy, the averages are obtained from (\ref{Eq:i-h}) and (\ref{Eq:PS_MI}) via Monte Carlo simulations. Since (\ref{Eq:i-h}) and (\ref{Eq:PS_MI}) are computed for an instantaneous channel, their numerical expectations are computed by averaging over channel realizations drawn from $\mathbf{h}_b$, $\mathbf{h}_f$, $\mathbf{h}_r$ for the best-port, fixed port, and random port selection strategy respectively.}

We benchmark the proposed optimized constellations against baseline APSK schemes reported in \cite{bayguzina2019asymmetric}. For fairness, the APSK constellations are generated with the same phase range $\delta$ as the optimized AQAM constellations, and all simulations are carried out under identical conditions. The results in Fig.~\ref{fig:rate-energy-region} demonstrate that the optimized constellations substantially outperform APSK, particularly under both best-port and fixed-port selection scenarios. In particular, the optimized constellations achieve a significantly wider RE-region. {It is worth noting that, although APSK employs the same phase range and average transmit energy as the optimized AQAM, its inferior performance stems from its fixed and very low PAPR (e.g., APSK exhibits a PAPR of 1), which inherently limits energy harvesting efficiency since rectifiers benefit from high-amplitude signal variations.} Furthermore, the integration of fluid antennas further enlarges the achievable RE-region—most prominently under best-port selection—clearly surpassing the baseline fixed-port (i.e., single-antenna) configuration.

\begin{figure}
\centering
\includegraphics[width= 0.7\linewidth]{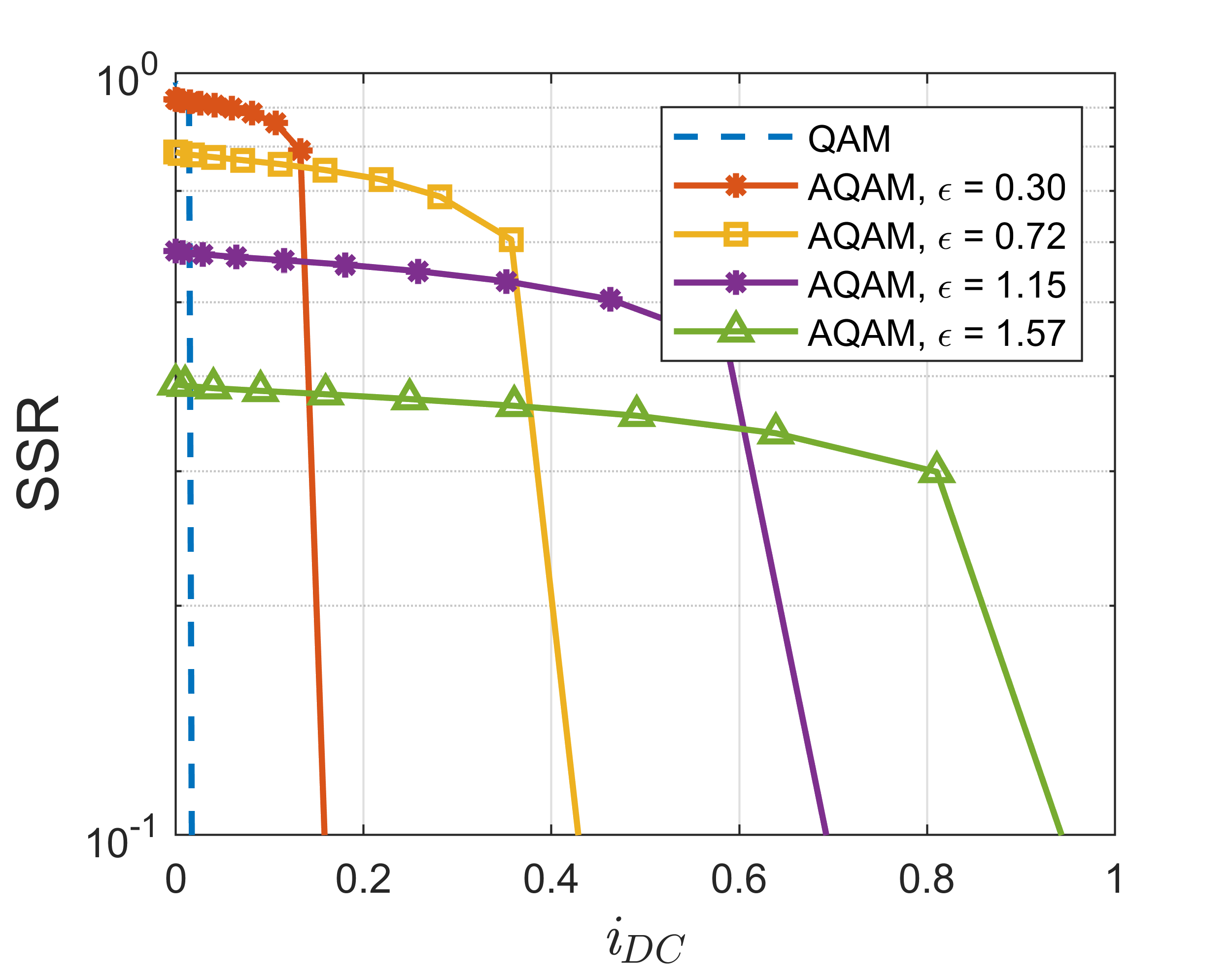} 
\caption{$\textit{i}_{DC}$ and SSR performance of optimized constellation schemes. The simulation parameters:  $M = 16$, $\rho = [0,1]$, $N = 100$  and $W = 0.5$.}
\label{fig:SSR-i_h}
\end{figure} 

\begin{figure}
    \centering    
    \includegraphics[width=0.8\linewidth]{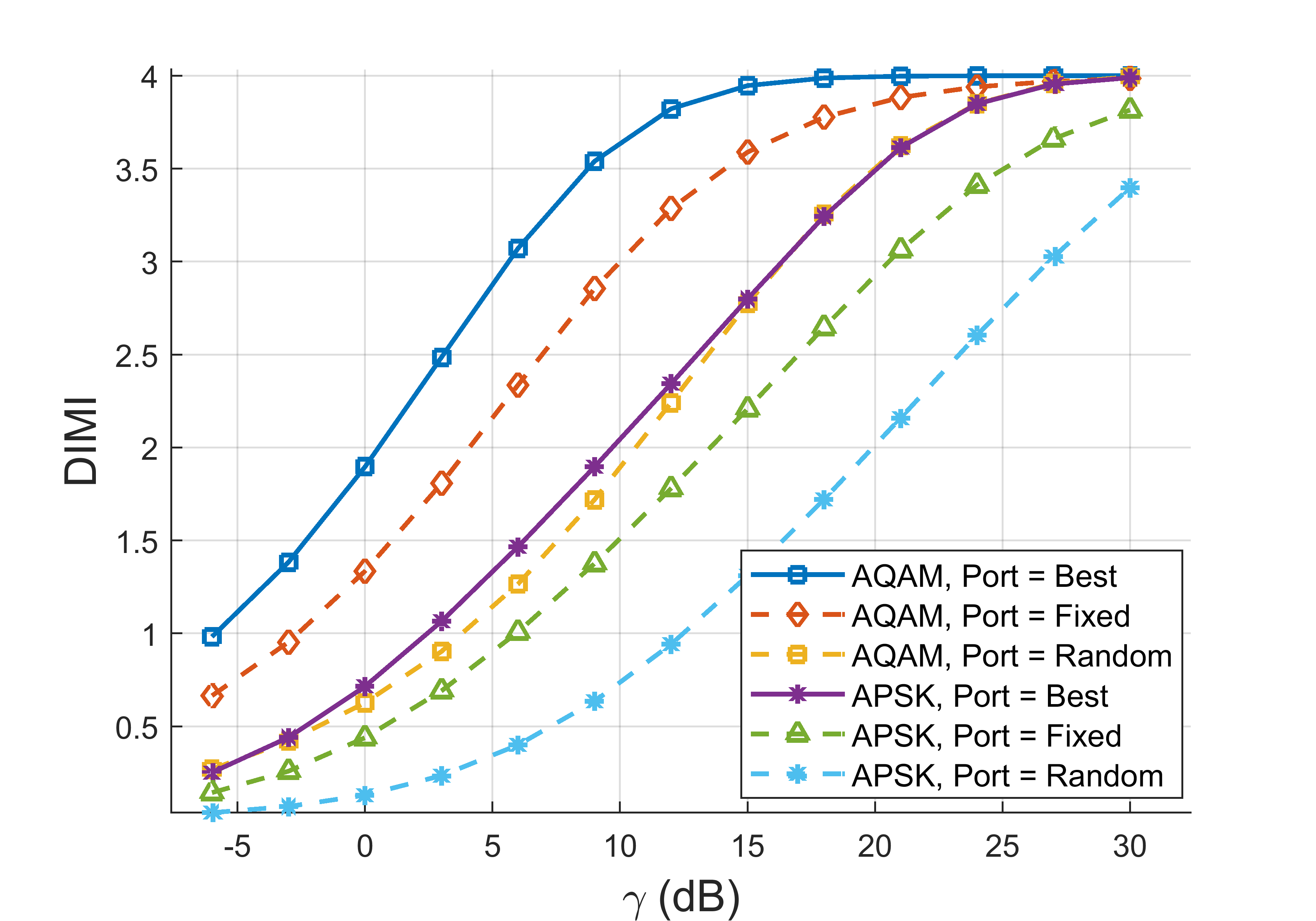}
    \caption{Impact of AQAM optimization on the DIMI. The simulation parameters are $M = 16$, $N = 100$, $\epsilon = 0.3.$}
    \label{fig:MI}
\end{figure}

We further investigate the impact of {SNR} on the DIMI for both the optimized constellations and the baseline APSK scheme. The average DIMI is computed via Monte Carlo simulation using (\ref{Eq:PS_MI}) across all three port selection strategies, with results shown in Fig. \ref{fig:MI}. The figure illustrates that APSK suffers from poor DIMI performance, primarily because the constellation point spacing becomes extremely small for narrow phase ranges, thus negatively affecting the DIMI. In contrast, the geometrically optimized constellations maintain larger inter-point distances by design, thereby significantly improving the DIMI and overall communication performance.

Finally, we evaluate, via Monte Carlo simulations, the trade-off between the symbol success rate (SSR) and the harvested current by varying the power-splitting factor $\rho$. {Note that the received SNR depends on $\rho$, whereas the harvested current $\textit{i}_{DC}$ depends on $1-\rho$. Consequently, as $\rho$ decreases from 1 to 0, the SNR is gradually reduced—resulting in a lower SSR—while the harvested current $\textit{i}_{DC}$ increases.
 To quantify this effect, we generate equally spaced values of $\rho$ in the range $[0,1]$ with a step size of $0.1$ using MATLAB.} For each $\rho$, the SSR and harvested current are computed through Monte Carlo simulations, where SSR is evaluated using the maximum likelihood detection rule. The resulting SSR–$\textbf{i}_{DC}$ trade-off curves in Fig.~\ref{fig:SSR-i_h} reveal that constellations optimized for lower $\epsilon$ values achieve higher SSR, while those optimized for larger $\epsilon$ values provide enhanced energy harvesting. These findings underscore the importance of adaptively selecting modulation schemes according to system requirements, thereby enabling an effective balance between information decoding and energy harvesting in SWIPT systems.

\section{Conclusion}\label{Conclusion}
This work presents a novel optimization framework for the design of geometrically shaped constellation schemes. By formulating and solving dual-objective RE-region optimization problem, we demonstrate that the optimized constellations significantly outperform conventional schemes like APSK in both mutual information and energy harvesting performance. Simulation results further highlight the advantage of fluid antennas in enhancing system flexibility and efficiency. {Future work will be to consider probabilistic constellation shaping in FA-assisted SWIPT systems.}

\section*{Acknowledgment}
{This work received funding from the European Research Council (ERC) under the European Union's Horizon 2020 research and innovation programme (Grant agreement No. 819819) and from the Smart Networks and Services Joint Undertaking (SNS JU) under the European Union's Horizon Europe research and innovation programme (Grant Agreement No 101192080).}

%% file: Tex_files/Appendix_A.tex
    \appendices
    \section{Proof of Proposition 1}
    \label{Prop1}
        
    To prove this preposition, we begin with (\ref{Eq:PS_MI}) and take the expectation with respect to $h_n$ as follows
         \vspace{-1mm}

        \footnotesize
        \begin{equation}
        \begin{split}
            I(X;Y)= m - \frac{1}{M}\mathbb{E}_{h_n}\Bigg\{\sum_{k=1}^M 
            \mathbb{E}_{N}   \left\{\log_2\sum_{i=1}^M e^{-f(h_n,x,\eta)}   \right\}\Bigg\},
        \end{split}
        \nonumber
        \end{equation}
        \normalsize

         \noindent where  $ f(h_n,x,\eta) = \frac{|\sqrt{\rho}h_m(x_k-x_i)+\eta|^2-|\eta|^2}{2 N_o}$ and  $m = \log_2(M)$. Now, to transform the expression into a more tractable objective function, it is necessary to remove the  expectations $\mathbb{E}_{h_n}\{.\}$ and $\mathbb{E}_{N}\{.\}$. To this end, we write $|\sqrt{\rho}h_n(x_k-x_i)+\eta|^2 = |\eta|^2 + \rho|h_n|^2|x_k-x_i|^2-2\sqrt{\rho}\Re(\eta|h_n|^2({x_k}-{x_i})^*)$) and simplify $ f(h_n,x,\eta)$ as
         \vspace{-1mm}

        \footnotesize
        \begin{equation}
             f(h_n,x,\eta) = \frac{|\sqrt{\rho}h_n(x_k-x_i)|^2+2\sqrt{\rho}\Re(\eta|h_n|^2({x_k}-{x_i})^*}{2N_o}.
            \nonumber
        \end{equation}
        \normalsize
        
        \noindent We now note that $\log_2\sum e^{-f(h_n,x,\eta)}$ has become a convex function of $\eta$ since $f(h_n,x,\eta)$ is an affine function of $\eta$. We exploit this convexity and take the expectation $\mathbb{E}_N\{.\}$ inside the $\log_2\sum \exp(.)$ by invoking Jensen's inequality $\mathbb{E}_X\{f(X) \}\geq f(\mathbb{E}_X\{X\})$ as follows,
        \vspace{-1mm}
        %\footnotesize
        \begin{equation}
            I(X;Y) \geq m - \frac{1}{M}\mathbb{E}_{h_n}\sum_{k=1}^M  \log_2\sum_{i=1}^M e^{\mathbb{E}_N\{-f(h_n,x,\eta)\}},
            \nonumber
        \end{equation}
        \normalsize     
        
        \noindent where $\mathbb{E}_N\{-f(h_n,x,\eta)\} = {-\rho|h_n|^2}|x_k-x_i|^2/ {2 N_o}$ as $\mathbb{E}_N\{2\sqrt{\rho}\Re(\eta|h_n|^2({x_k}-{x_i})^*)\} = 0$. The lower bound can now be expressed as,
        \vspace{-1mm}

        \footnotesize
        \begin{equation}
            I_{LB}(X;Y) = m - \frac{1}{M}\mathbb{E}_{h_n}\Bigg\{\sum_{k=1}^M  \log_2\sum_{i=1}^M 
            e^{\frac{\rho|h_n|^2}{2  N_o}|x_k-x_i|^2 ) }\Bigg\}.            
            \nonumber
        \end{equation}
        \normalsize
        
        \noindent To simplify the above equation further, we also take $\mathbb{E}_{h_n}$ inside the $\exp(.)$ function as follows,
        
        \footnotesize
        \begin{equation}\label{Eq:PS_MI_LB}
            \tilde{I}_{LB}(X;Y) \approx  m - \frac{1}{M}\sum_{k=1}^M  \log_2\sum_{i=1}^M 
            e^{-\frac{\rho\mathbb{E}\{|h_n|^2\}}{2 N_o}|x_k-x_i|^2 }.             
            \nonumber
        \end{equation}
        \normalsize
        
        Note that, we can initially compute $\mathbb{E}\{|h_m|^2\}$ and feed it to the objective function, eliminating the need of the Monte Carlo evaluation of the objective function during the optimization process.